\documentstyle[twocolumn,aps,psfig]{revtex}
\input psfig.sty

\newcommand{\be}{\begin{equation}}
\newcommand{\ee}{\end{equation}}

\begin{document}
\title{Stochastic Time-Dependent Hartree-Fock for heavy-ion collisions in 
the Fermi energy domain}
\author{E. Suraud
\thanks{Membre de l'Institut Universitaire de France}}
\address{Laboratoire de Physique Quantique, 
Universit\'e Paul Sabatier,
118 route de Narbonne,
31062 Toulouse Cedex, France}
\author{P. G. Reinhard}
\address{Institut f\"ur Theoretische Physik, Universit\"at Erlangen, 
Staudst. 7, D8520 Erlangen, Germany}
\date{\today}
\maketitle

\begin{abstract}
We present the first realization of Stochastic TDHF, a theory which
goes beyond pure mean-field dynamics, embracing dissipation as well as
fluctuations. Applications to heavy-ion collisions in the Fermi energy
domain are given and analyzed in terms of dissipative features and
Intermediate Mass Fragment production.
\end{abstract}


Since its early introduction in quantum dynamics Time-Dependent
Hartree-Fock (TDHF) \cite{dirac} is the basic selfconsistent theory
for many dynamical processes in physics and chemistry. In nuclear
physics, it has found since long widespread and fruitful applications
in collective motion and heavy-ion reactions
\cite{RevNeg,Honneff,Klakow,BerFel}. Recently, a new field of
applications to the nonlinear electron dynamics in metal clusters has
opened up \cite{TDLDAP1,ISSPICBer,TDLDAP2} which  brings up
challenging questions in connection with new experimental
investigations \cite{chand,haberland,barat,wohrer}.
 
Mere TDHF is nevertheless too restricted in its degrees-of-freedom if
one goes beyond the regime of low-energy (collective) dynamics. For
example, TDHF becomes insufficient for the description of central
heavy-ion collisions in the Fermi energy domain \cite{hotn}.  But it
is just these collisions which have unravelled a bunch of exciting
phenomena, particularly from the generation of experiments using $4
\pi$ multidetectors \cite{imf}, and there exists up to now no
theoretical tool which is definitively validated in this field
\cite{esrw}.

Extensions of TDHF to include dynamical corelations are formally and
computationally feasible \cite{Cusson,BalVen,Koonin}. But these
extensions still remain restricted to the low-energy regime, where only
few channels are active in the dynamics. Statistical approaches are
more appropriate for higher energies and there exist attempts to
complement the TDHF equation by a collision term
\cite{DavWon,Koehler}. This raises, however, considerable technical
difficulties due to the large number of rarely occupied orbitals
needed in such an approach. Moreover, such a collisionally extended
TDHF is still bound to one mean path and grossly underestimates the large
statistical fluctuations of the mean field.

The semiclassical variant of collisional TDHF, the Boltzmann - Uehling
- Uhlenbeck (BUU) approach, has been more successful in describing
average trends in medium-energy heavy ion collisions
\cite{BerGup,MosCas}.  But it suffers also from insufficiencies,
particularly when considering experiments in which fluctuations are
playing a crucial role such as in the formation of Intermediate Mass
Fragments (IMF's) \cite{imf} or in the production of rare particles
well below threshold \cite{part}.  The Boltzmann-Langevin equation
(BLE) has thus been introduced as 
an extension of BUU which incorporates (possibly large)
dynamical fluctuations by switching to a full ensemble description.
It should allow a pertinent description of variances and widths of
observables \cite{RevAbe}. But huge computational (and even some
formal) difficulties hinder up to now large scale applications of
these BLE approaches.

A related theory for generating and propagating an ensemble of quantum
states is Stochastic Time-Dependent Hartree-Fock (STDHF). It goes back
to the basic quantal mean field dynamics and treats dynamical
correlations in Markovian approximation by occasional jumps from one
TDHF trajectory to another one, thus generating an ensemble of TDHF
trajectories in the course of time \cite{STDHF1}. It is interesting to
note that similar trajectory hopping techniques have also succesfully
been applied to electron dynamics in small molecules \cite{hopp}. From a formal point of
view, STDHF is the basic theory from which collisional TDHF, BUU and
BLE can be derived by appropriate subsamplings \cite{STDHF1}. And
technically, it offers a particularly efficient scheme to evaluate at
once collision as well as fluctuation terms \cite{RevAbe} by
means of trajectory hopping. The ensemble of pure Slater states is a
more efficient way of handling thermalization because it circumvents
the complication to carry an extra load of nearly unused single
particle wavefunctions.  Furthermore, its quantal nature avoids the
many (formal and numerical) difficulties encountered in a proper
justification \cite{husimi,esrw} and simulation of semi-classical
kinetic equations \cite{pges1} in the context of heavy-ion collisions.

The key issue in the STDHF scheme is an efficient selection of the
jumps. Strategies have been proposed, based on the principle of
maximum overlap with the corresponding correlated state and evaluated
practically with linear response techniques \cite{Aussois,Brioni}. But
before attacking the problem with these elaborate (and costly)
optimization principles, it is worthwhile to investigate the
capabilities of STDHF as such, by using an intuitive approach to the
jumps. Our procedure is inspired by BUU and BLE, where the 
(hardly tractable) 5-fold local Boltzmann collision integral 
is replaced in a robust manner by sampled collisions of test particles, 
simply parametrized by one parameter, a total effective nucleon nucleon 
cross section $\sigma$ \cite{welke}.  It is the
aim of the present letter to propose such a simple, intuitive, and
efficient scheme for generating the ensemble of STDHF, and to apply it
to a realistic test case, collisions of $^{16}$O on $^{16}$O at medium
energies, where pure TDHF only shows transparency, while dissipation
should lead to incomplete fusion, fission or fragmentation
\cite{hotn}.

First, let us recall briefly the evaluation of the collision term in
BUU \cite{BerGup,welke}. The total cross section $\sigma$ is rescaled
to an effective cross section for the test particles $\sigma_{\rm
eff}=\nu\sigma$, where $\nu$ is the number of particles per
testparticle \cite{welke}. For all pairs of test particles "(1,2)" one
checks whether "1" and "2" will approach each other closer than a
scattering length $d_\sigma=\sqrt{\sigma_{\rm eff}/\pi}$, within the
next time step. If yes, "1" and "2" collide, i.e.  their momenta are
changed $(p_1,p_2)\longrightarrow(p'_1,p'_2)$, while preserving total
momentum and total kinetic energy. Finally, the Pauli-blocking factors
$(1-f'_1)(1-f'_2)$ are computed at the new phase-space points
$(r_1,p'_1)$ and $(r_2,p'_2)$ and sampled by Monte-Carlo techniques.

A TDHF state $|\Phi\rangle$ consists of single particle wavefunctions
$\varphi_\alpha(r,t)$ where each one represents one physical particle
and is spread over all space, carrying a local density distribution
$\rho_\alpha(r,t)$ and a local momentum distribution 
$p_\alpha(r,t)=j_\alpha(r,t)/\rho_\alpha(r,t)$. 
Collisions, on the other hand, are
local. We thus split our space grid into "test volumes" 
${\cal V}^{(i)}$ which are going to play the role of the test particles in
BUU. A spatial filtering function ${\cal G}^{(i)}(r)$ is associated
with this test volume and we define ${\cal V}^{(i)}$-test volume
integration as 
$\int_{{\cal V}^{(i)}} d{\bf r} ...=\int d{\bf r}{\cal G}^{(i)}(r)...$. 
For a given ${\cal V}^{(i)}$, we hence evaluate the "local" average positions 
$\bar{r}^{(i)}_\alpha=\int_{{\cal V}^{(i)}}{\bf r}\rho_{\alpha} d{\bf r} /
N^{(i)}_{\alpha}$ 
and momenta 
$\bar{p}^{(i)}_\alpha=\int_{{\cal V}^{(i)}}j_{\alpha}({\bf r}) 
d{\bf r} / N^{(i)}_{\alpha}$, 
where
$N^{(i)}_{\alpha}=\int_{{\cal V}^{(i)}}\rho_\alpha d{\bf r}$.  
These
positions and momenta are then handled in exactly the same manner as
in BUU. The only modification is that the particle content $\nu$ now
becomes a dynamical quantity to be evaluated as the average fraction
of the two scattering particles $\alpha$ and $\beta$ in ${\cal
V}^{(i)}$, i.e. $\nu=(N^{(i)}_\alpha N^{(i)}_\beta)^{1/2}$. 

We thus obtain a jump of the associated momenta
$(\bar{p}^{(i)}_\alpha,\bar{p}^{(i)}_\beta) \rightarrow
(\bar{p}'^{(i)}_\alpha,\bar{p}'^{(i)}_\beta)$, obeying local
conservation of local momentum and kinitic energy as
\begin{eqnarray}
  \bar{p}^{(i)}_\alpha+\bar{p}^{(i)}_\beta
  &=&
  \bar{p}'^{(i)}_\alpha+\bar{p}'^{(i)}_\beta
  \quad,
\\
  (\bar{p}^{(i)}_\alpha)^2+(\bar{p}^{(i)}_\beta)^2
  &=&
  (\bar{p}'^{(i)}_\alpha)^2+(\bar{p}'^{(i)}_\beta)^2
  \quad.
\end{eqnarray}
We need now to modify the phase profile of the single particle
wavefunctions $\varphi_\alpha$ and $\varphi_\beta$ such that local
average over ${\cal V}^{(i)}$ produces the new momenta
$(\bar{p}'^{(i)}_\alpha,\bar{p}'^{(i)}_\beta)$. This is achieved by
the transformation 
\begin{equation}
  \varphi'_\alpha(r,t)
  =
  \exp{\left(-i(\bar{p}'_\alpha-\bar{p}_\alpha)r{\cal G}^{(i)}(r)\right)}
  \varphi_\alpha(r,t)
\label{eq:jumpwf}
\end{equation}
and similarly for $\varphi_\beta$. 

As in BUU, the collision may end up in a region of phase space where
particles are already present. This is measured by the overlap of the
old $|\Phi\rangle$ and new $|\Phi'\rangle$ TDHF states.  The Pauli
blocking factor is the amount of still available space, which is then
simply $1-\left|\langle\Phi'|\Phi\rangle\right|^2$. This blocking
factor is sampled by Monte-Carlo techniques. In contrary to the
classical BUU scheme we need here two more steps for a final
clean-up. First, the new single particle wavefunctions
$\varphi'_\alpha$ and $\varphi'_\beta$ are each orthogonalized on all
other $\varphi_\gamma$. And second, we need to restore the original
value of the total energy because we find a slight modification of the
total energy (although most of the energy matching has been accounted
for by conserving the "classical" kinetic energy in the collision).
This is achieved by energy rescaling with an iterative imaginary time
step $|\Phi'\rangle\longrightarrow(1-\delta\hat{h})^n|\Phi'\rangle$
where $\hat{h}$ is the actual mean field Hamiltonian and $\delta$ a
(small) shift parameter to be chosen such that total energy is
recovered in a few iterations $n \sim 4-5$.  This altogether
accomplishes the stochastic jump of the STDHF trajectories.

The scheme can be simplified for the initial stages of a collision of
two small nuclei. The single particle wavefunctions in each one of the
colliding nuclei are initially well localized, such that we can skip
the subdivision into test volumes and treat the collision with respect
to the straightforward averages $\bar{r}_\alpha$ and $\bar{p}_\alpha$
of the single particle wavefunctions. This amounts to letting the
space-filtering factor ${\cal G}^{(i)}\longrightarrow 1$ and the
trivial renormalization factor
$\nu=(N^{(i)}_{\alpha}N^{(i)}_\beta)^{1/2}\longrightarrow1$. This
simplification is surely applicable for the initial stages of the collision
where most of the statistical ensemble is built up and where most
dissipation takes place \cite{RevAbe}.  It may overestimate collision
rates after the compound stage. But this is mostly a quantitative
problem. The description will remain pertinent at least at a
qualitative level.

For a first exploration, we have implemented the above scheme in its
simplified version (using global averages rather than local ones) into a
3D nuclear TDHF code without any symmetry restriction. The mean field
Hamiltonian is generated from a Skyrme force with a density-dependent
zero-range interaction, as used in  BUU studies \cite{pges1}. 
As a test case, we consider  $^{16}$O on $^{16}$O collisions.

\begin{figure}[t]
\psfig{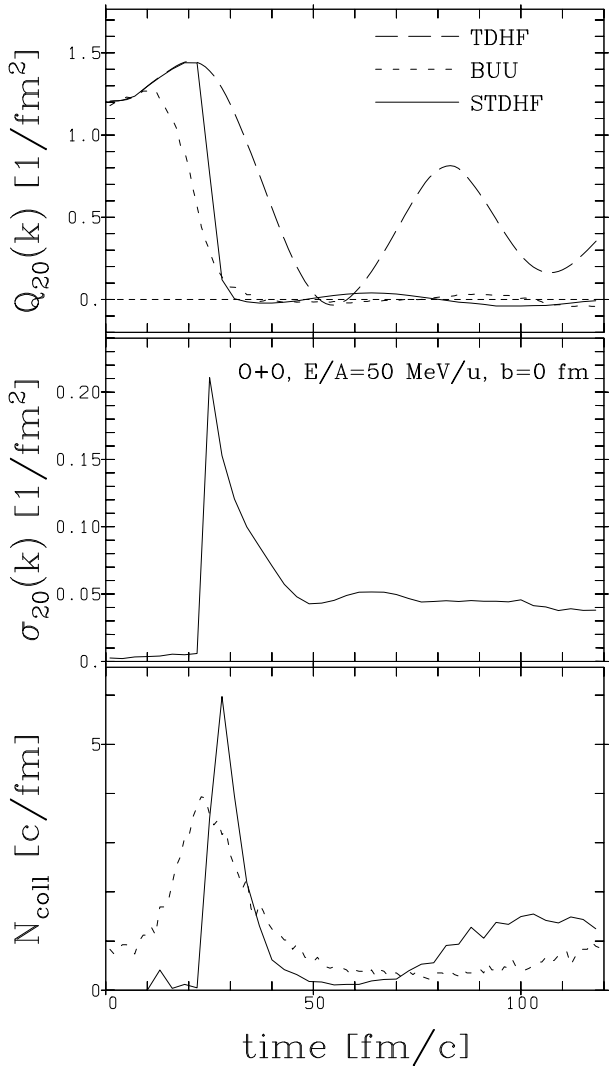}
\vspace*{4mm}
\caption{
Time evolution of a collision $^{16}$O+$^{16}$O with impact parameter
$b=0\,{\rm fm}$ at $50\,{\rm MeV/u}$ beam energyi (STDHF: ensemble of 100 events,
BUU and TDHF: 1 (average) calculation). 
Uppermost block: the
quadrupole moment $Q_{20}=\langle k^2 Y_{20}\rangle$. 
Second block: variance of $Q_{20}$ in the STDHF ensemble.
Third block: number of successful 2-body collisions per fm/c
in STDHF  and BUU.
}
\label{fig1}
\end{figure}

Figure \ref{fig1} shows the results for a head-on collision
$^{16}$O+$^{16}$O at $50\,{\rm MeV/u}$ beam energy.  The
STDHF scheme has been carried up to an ensemble of 100 trajectories.
The upper panel shows the time evolution of the ensemble average momentum
quadrupole moment $Q_{20}=\langle k^2 Y_{20}\rangle$. It is known to
provide a robust indicator of the dissipative phase leading to the
formation of an excited compound nucleus (distinguished by a small
deformation) \cite{bauer,RevAbe}.  A TDHF and a BUU result are also
plotted for comparison.  The TDHF calculation leads in this case to a
perfect transparency which is reflected by typical mean-field
oscillations of $Q_{20}$. In contrast, the STDHF calculation, 
as BUU, leads to the formation of a compound nucleus as can be seen
from the damping of $Q_{20}$ to 0.  The damping rate $\tau_c$ in
STDHF, however, is somewhat larger than in BUU.  This rate may be
further analysed by considering the number of succesful two-body
collisions $n_c$ as a function of time, see lower panel of Figure
\ref{fig1}. Two-body collisions are peaked around $t\!=\!30\,{\rm
fm/c}$, which roughly corresponds to the time of maximum overlap of
the two colliding nuclei. The BUU calculation provides a smaller
peak value of $n_c$, but a broader peak, such that, after all, the
integrated number of two-body collisions during the overlap phase is
comparable to the result of STDHF.  
It is interesting to note here the
difference between BUU and STDHF, {\em before} overlap. While in STDHF
Pauli blocking is perfectly effective, leading to exactly {\em no} two
body collisions ($n_c = 0$), there remains a residual background of
2-body collisions in the BUU case, due to the approximate treatment of
Pauli blocking and to the fact that the test particle sampling only
provides approximate ground states, even with large numbers of test
particles per nucleon. Finally, the middle panel of Figure \ref{fig1}
displays the variance $\sigma_{20}$ associated to the quadrupole
moment $Q_{20}$, as calculated from the ensemble of computed
events. Note that $\sigma_{20}$ exhibits a marked bump at the time of
maximum overlap.  Such a bump in variance is characteristic for a
transient regime as demonstrated in BLE-based calculations
\cite{RevAbe,transient}.

\begin{figure}
\psfig{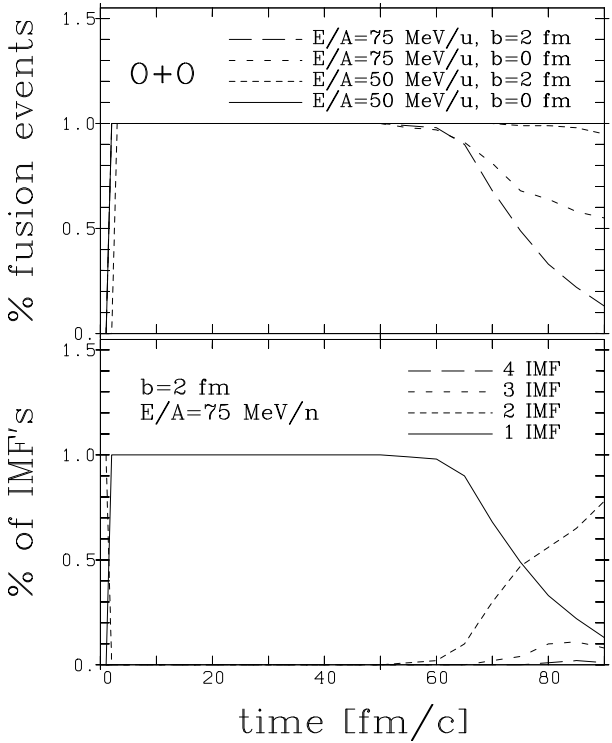}
\vspace*{4mm}
\caption{
Time evolution of the fraction of events (100 events ensembles for 
each reaction)
leading to incomplete fusion 
(upper panel) and to fission of fragmentation (lower panel) for 
$^{16}$O+$^{16}$O collisions with impact parameters
$b=0\,{\rm and \,}2\,{\rm fm}$ 
at $50\,{\rm and \,}75 \,{\rm MeV/u}$ beam energy.
}
\label{fig2}
\end{figure}

As stressed above, a major advantage of STDHF is the fact that it
provides at once an ensemble of TDHF trajectories, which contain the
proper (possibly large) fluctuations of the mean-field.  As a first
illustration of the capabilities of STDHF, we analyze the emerging
ensemble in terms of the formation of IMF's.  The IMF identification
is done with a percolation-type algorithm applied to the density of
matter inside the computing box, as in \cite{elba}. This analysis
allows to evaluate, mass, charge, velocity,...  of the IMF's. The
upper panel of Figure \ref{fig2} shows the fraction $f_1$ of events
leading to incomplete fusion (1 fragment) as function of time. Results
from four different $^{16}O + ^{16}O$ collisions at beam energies 50
and 75 MeV/u and impact parameters $b\!=\!0$ and $2$ fm are shown.
Just after touching ($t\!\sim\!2\,{\rm fm/c}$), $f_1$ goes to 1 which
corresponds to the overlap phase.  In a later stage
($t\!\ge\!40-50\,{\rm fm/c}$) the system evolves to either incomplete
fusion ($f_1\!=\!1$), fission (2 IMF's), or
fragmentation (3 or more IMF's). At 50 MeV/u, both
impact parameters lead to almost $100 \%$ incomplete fusion (average
mass $<A>\!\sim\!22\!\pm\!1$ and charge $<Z>\!\sim\!11\!\pm\!1$). Note
that isospin symmetry (N=Z) is preserved here, without being enforced.
At 75 MeV/u, fission and even fragmentation mechanisms take over.  The
detailed division into various fragments is shown for the most "fragmenting" 
case in the lower panel of Figure \ref{fig2}.  Fission seems still to
dominate. But there is a sizeable amount of events with three
fragments and occasionally more than that.
 
It should finally be noted that the computational effort involved in
one STDHF event is comparable to the effort required for performing an
acceptable BUU calculation. This has to be put in perspective with the
differences in the physical contents of the 2 approaches (quantal
versus semi-classical, full fluctuations versus average description).

We have presented a first application of STDHF as a propagation scheme
which is capable of describing collisions and fluctuations beyond a
mean-field approach by generating an ensemble of TDHF trajectories
through occasinal jumps.  An intuitively motivated and well manageable
jump algorithm was developed in close analogy to the semiclassical
cousin of collisional TDHF, the BUU equation. The first results for
collisions $^{16}$O+$^{16}$O are very encouraging.  We see a
physically reasonable damping of collective motion leading to
formation of a compound nucleus. This proves that STDHF is an efficient
way of sampling a collision term in the case of quantal
propagation. Beyond that, STDHF generates at the same time an ensemble
of trajectories with the capability of embracing large 
mean-field fluctuations, which corresponds to the Boltzmann-Langevin
treatment in the semiclassical domain.  This also allows a reliable
description also of variances. The results of our test case show
nicely the growth of the variance related to strong dissipation.
Moreover, these results have been obtained with much less expense 
but more physics than needed for a comparable BLE calculation.

\bigskip

Acknowledgements: The authors thank the french german exchange program 
PROCOPE (grant number 95073) and Institut Universitaire 
de France for financial support. F. Calvayrac is  
thanked for providing a version of his 
3-D TDLDA code and P. Bonche is acknowledeged for discussions in the final
stages of the work.

\end{document}